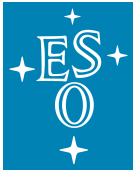

ESO Expanding Horizons White Paper

# Electromagnetic characterisation of millihertz gravitational-wave sources in the Milky Way

**Contact authors**
**James Munday** University of Warwick, *Coventry, United Kingdom;* james.munday98@gmail.com
**Valeriya Korol** Max Planck Institute for Astrophysics, *Garching, Germany;* korol@mpa-garching.mpg.de
**Camilla Danielski** University of Valencia, *Valencia, Spain;* camilla.danielski@uv.es

**Proposing team**
**Na'ama Hallakoun** Weizmann Institute of Science, *Rehovot, Israel*
**Astrid Lamberts** Observatoire de la Côte d'Azur, *Nice, France*
**Gijs Nelemans** Radboud University, *Nijmegen, The Netherlands*
**Anna Pala** European Southern Observatory, *Garching, Germany*
**Steven Parsons** University of Sheffield, *Sheffield, United Kingdom*
**Ingrid Pelisoli** University of Warwick, *Coventry, United Kingdom*
**Alberto Rebassa Mansergas** Polytechnic University of Catalonia, *Barcelona, Spain*
**Jan van Roestel** Institute of Science and Technology Austria, *Klosterneuburg, Austria*

**Supporters (109 astronomers endorsed this white paper in total)**
Borja Anguiano (Centro de Estudios de Física del Cosmos de Aragón, Spain), K. G. Arun (Chennai Mathematical Institute, India), Maria Babiuc Hamilton (Marshall University, USA), Carles Badenes (University of Pittsburgh, USA), John G. Baker (Université de Toulouse, France), Manuel Barrientos (University of Oklahoma, USA), Enrico Barausse (SISSA, Italy), Alexander Bobrick (Monash University, Australia), Elisa Bortolas (INAF, Italy), Tamara Bogdanovic (Georgia Institute of Technology, USA), Alexander Bonilla Rivera (Universidade Federal Fluminense, Brazil), Martin A. Bourne (University of Hertfordshire, UK), Warren R. Brown (CfA, USA), Tomasz Bulik (University of Warsaw, Poland), Riccardo Buscicchio (University of Milano-Bicocca, Italy), Mario Cadelano (University of Bologna, Italy), Ana Caramete (Institute of Space Science, Romania), Laurentiu-Ioan Caramete (Institute of Space Science, Romania), Amodio Carleo (INAF, Italy), Sylvain Chaty (Université Paris Cité, France), Michael W. Coughlin (University of Minnesota, USA), Stefano Covino (INAF, Italy), Alexander W. Criswell (Vanderbilt University, USA), Gergely Dálya (L2IT, France), Benjamin L. Davis (New York University Abu Dhabi, United Arab Emirates), Harry Dawson (University of Potsdam, Germany), Pratika Dayal (CITA, Canada), Kyriakos Destounis (Universidade de Lisboa, Portugal), Tiziana Di Salvo (University of Palermo, Italy), Thomas Donlon (University of Alabama in Huntsville, USA), Matti Dorsch (University of Potsdam, Germany), Hannah Dykaar (McGill University, Canada), Kareem El-Badry (Caltech, USA), Chris Fryer (Los Alamos National Laboratory, USA), Ioannis D. Gialamas (Laboratory of High Energy and Computational Physics, Estonia), Poshak Gandhi (University of Southampton, UK), Silvia Gasparotto (CERN, Switzerland), Sebastien Guillot (IRAP, France), Leonid Gurvits (TU Delft, The Netherlands), Konstantinos Nektarios Gourgouliatos (University of Patras, Greece), Ulrich Heber (University of Erlangen–Nürnberg, Germany), Joerg Hennig (University of Otago, New Zealand), JJ Hermes (Boston University, USA) and additional endorsing scientists worldwide (full list available here).

## Context

Gravitational-wave astronomy will transform in the 2030s as space-based observatories access new frequency regimes. The European Space Agency's flagship Laser Interferometer Space Antenna (LISA)—adopted in 2024 and planned for launch in 2035 (*Colpi et al. 2024*)—alongside the Chinese missions TianQin (*Luo et al. 2016*) and Taiji (*Ruan et al. 2018*) will open the millihertz (0.1–100 mHz) window. Looking ahead to the 2040s–2050s, multiple mission concepts have been proposed to open the decihertz and microhertz bands, including lunar-based detectors, further extending coverage of the gravitational-wave spectrum (e.g. *Arca Sedda et al. 2020; Sesana et al. 2021; Ajith et al. 2024*). Collectively, these projects will enable studies of astrophysical and cosmological phenomena that are not directly accessible by other means.

The millihertz band is densely populated by continuous gravitational-wave signals from Galactic compact binaries, dominated by double white dwarfs (DWDs; binaries of two white dwarfs) with contributions from systems containing neutron stars and black holes (*Amaro-Seoane et al. 2023*). As these binaries inspiral due to gravitational-wave radiation, they can reach contact and begin mass transfer in the millihertz band. Gravitational-wave detectors like LISA will survey such compact binaries across the Milky Way, yielding samples numbering in the tens of thousands, with essentially complete sensitivity to orbital periods shorter than ~10-20 min (e.g. *Lamberts et al. 2019*). Assessing the nature of the binary components—and deriving masses, temperatures and compositions—requires systematic electromagnetic characterisation that breaks gravitational-wave degeneracies and enables full atmospheric and orbital solutions. At present, no dedicated facility or coordinated survey is planned to deliver electromagnetic follow-up at the scale necessary to maximise the science return of the millihertz gravitational-wave data; this white paper discusses the need and requirements of such a capability.

## Science questions

**I What is the nature and demographics of Galactic millihertz gravitational-wave sources?**

Confirming whether gravitational-wave detections are DWDs or contain neutron star/black hole components is not straightforward from gravitational-wave data alone. Yet, it is a necessary prerequisite to constrain binary evolution models and to study them in a broader context of the Milky Way's formation and evolution. For most detectable sources (~85%), the gravitational-wave frequency will be effectively constant over the mission baseline. In these cases, the nature of a source can only be inferred probabilistically from gravitational-wave observables, and it is generally not possible to completely discriminate between a nearby DWD and a more distant Galactic black hole binary. For a minority of sources (~15%), the first frequency derivative—the so-called gravitational-wave chirp—will be measurable, enabling an estimate of the chirp mass (a combination of the two component masses) and the distance. While this provides some leverage for separating different binary types by mass scale, it does not yield the individual component masses necessary to pin down the nature of the binary. Figure 1 illustrates that theoretical expectations yield several hundred (a few thousand) DWD counterparts brighter than *r*<21 mag (*r*<27 mag). ***A dedicated spectroscopic survey for LISA electromagnetic***

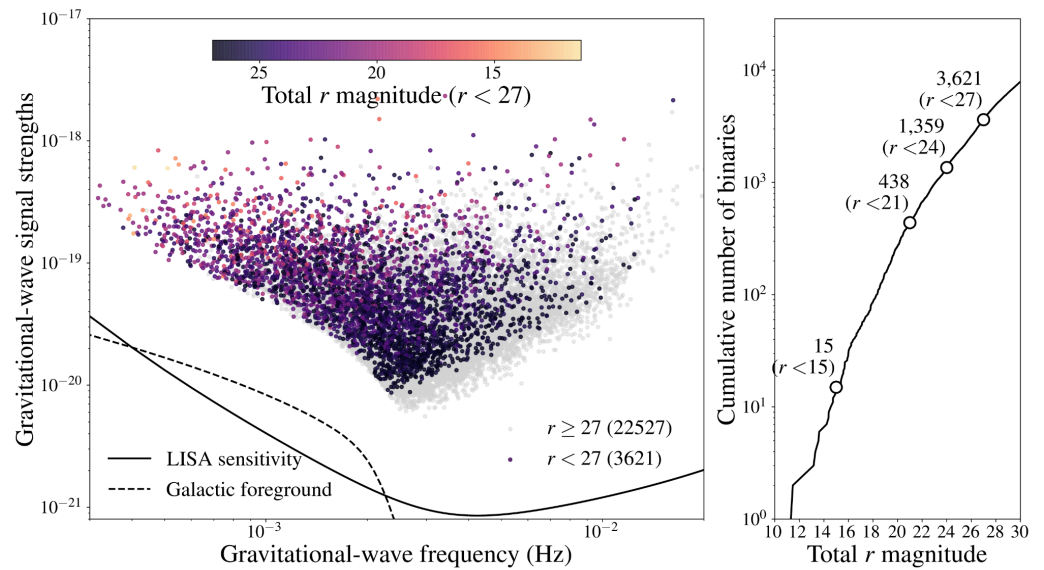

*Figure 1: LISA-detectable Galactic DWD binaries, coloured by total r-band magnitude; systems with r<27 are shown in colour, fainter ones in grey. Mock catalogue based on Korol et al. 2017.*

*counterparts is required to confirm DWDs, while an undetected flux from one of the two stars is a diagnostic for white dwarf plus neutron star or black hole binaries.*

**II What is the Galactic type Ia supernova merger rate due to DWDs?**

While type Ia supernovae are abundant in transient surveys and serve both as cosmic standard candles and as major contributors to the enrichment of heavy elements, their dominant progenitor channel remains debated. DWD binaries have long been among the most favoured type Ia progenitors (see *Ruiter & Seitenzahl 2025*). However, only 3% of the type Ia supernova rate in the Milky Way can currently be accounted for observationally (*Pelisoli et al 2021*; *Munday et al. 2025*), significantly lower than expected from synthetic predictions (e.g. *Rebassa-Mansergas et al. 2019*), although the total DWD merger rate is a few times higher than the type Ia supernova rate (*Maoz et al. 2018*). LISA, for example, will deliver a few thousand DWDs with chirp-mass measured to better than 30%, constraining the DWD merger rate to ~4-9% precision (*Korol et al. 2024*). Type Ia explosion models show that explosion energetics depend sensitively on the component masses (e.g. *Pakmor et al. 2012; Boos et al. 2024*). Thus, a chirp-mass measurement alone is insufficient to recover the individual component masses as it only provides a lower bound on the primary and leaves the companion weakly constrained, preventing us from identifying which LISA-selected DWDs can produce type Ia supernovae. To understand the full diversity of type Ia explosions and to connect DWD progenitors with explosion models, *radial velocity curves of gravitational-wave-selected DWDs are therefore required to obtain accurate masses for both stars.*

**III What fraction of gravitational-wave sources evolve into interacting binaries rather than merging?**

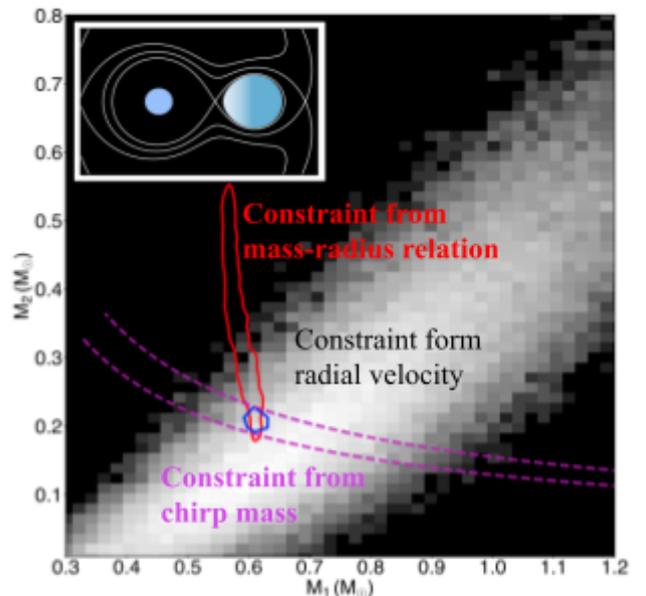

Figure 2: Constraints on component masses in ZTFJ1539+5027; Adopted from Burdge et al. 2019

As binaries reach orbital periods ≲10 min (gravitational-wave frequency ≳3 mHz), the interplay of tides, gravitational-wave losses and mass transfer determines the fate of the system: prompt merger/thermonuclear explosion (e.g. a type Ia supernova) or transition to a stably interacting binary (e.g. AM CVn). Which outcome occurs depends on the mass ratio, the donor's response to mass loss, and, crucially, the rate at which the white dwarfs' spins resynchronise with the orbit after spin-up by mass transfer. As ultra-short-period systems are very hard to find electromagnetically (only a handful are known), the merger-versus-stable-interaction branching fraction is essentially unconstrained—yet these binaries are ideal gravitational-wave sources. For ultra-short-period systems, LISA will measure chirps, allowing a direct estimate of the branching ratio between inspiralling binaries (positive chirps) and out-spiralling, stably interacting binaries (negative chirps). With hundreds expected in each channel, the ratio can be constrained to better than ~10% (*Colpi et al. 2024*). However, the chirp measurement blends effects of gravitational-wave radiation reaction, tides, and mass transfer, so gravitational-wave data alone cannot apportion their contributions. Only joint electromagnetic and gravitational-wave observations (Figure 2) can disentangle these effects and deliver a robust mass–period distribution and stability boundary—essential inputs to theoretical models. *Independent electromagnetic measurements combining light curve, radial velocity curve and orbital trajectory timing are required to constrain the individual white dwarf masses*.

**IV What is the occurrence and nature of circumbinary companions to gravitational-wave sources?**

Population studies indicate that a substantial fraction of millihertz Galactic binaries may have been influenced by a tertiary at some stage of their evolution (*Rajamuthukumar et al. 2025*). Theory likewise suggests that sub-stellar tertiaries around DWDs are plausible: planets around close binaries can preferentially survive host evolution (*Kostov et al 2016*), a large fraction (≈23–32%) of giant planets may persist to the DWD stage (*Columba et al 2023*), and fallback discs after the final common envelope can form close-in sub-Neptunes to giants within LISA's reach (*Ledda et al. 2023*). A tertiary induces a perturbation of the center of mass of the inner binary, imprinting a periodic Doppler modulation on the gravitational-wave signal (e.g. *Robson et al. 2018*). For outer-orbit periods shorter than the observing baseline, companions down to roughly Jupiter mass are detectable in gravitational-wave data (*Seto 2008*). Gravitational-wave dedicated data analysis can flag and partially characterise candidate tertiaries (*Katz et al. 2022*). ***Follow-up electromagnetic observations are required to confirm and fully characterise the nature of the companion in terms of periods, masses (of the three components), and inclination of the orbits.*** Depending on the electromagnetic method used, the companion parameters that can be retrieved can vary (see Tamanini & Danielski 2019 for details on the different electromagnetic synergies).

## Observing strategy, technology development and data handling requirements

Today, wide-field optical surveys (e.g. the Zwicky Transient Facility) discover compact binaries via blind, all-sky period searches. Low amplitudes and photometric errors comparable to the variability produce numerous period aliases, making many low-inclination systems and 30–90 min DWDs effectively unrecoverable. LISA will change this paradigm by pre-selecting fields and providing precise orbital periods, enabling the detection of fainter photometric variability and far more reliable counterpart identification in the 2040s. LISA localises Galactic sources through annual Doppler phase and a time-varying antenna response, with localisation precision improving with observing time. Galactic membership is inferred from the LISA distance, which is constrained by the measured strain amplitude together with the gravitational-wave chirp. After accounting for line of sight extinction and the Galactic locations of a mock sample, synthetic populations predict 438 sources with $r<21$ mag and 1359 $r<24$ mag sources (Figure 1). Typical LISA localisations are $\sim$1–10 deg$^2$ and include precise orbital periods. To identify electromagnetic counterparts, blue, white dwarf-like targets can be preselected through deep imaging (e.g. LSST, Pan-STARRS) and quasars rejected via (near-)zero proper motions. Because many candidates will remain per field, optical light curves will be phase-folded on the LISA period to isolate consistent variables. For orbital periods <1 hour, >0.05-0.1 mag variability occurs for inclinations above $\gtrsim$60-70 degrees (~30% of the time), yielding ~408 counterparts at $r<24$ mag that require spectroscopic confirmation.

**Target I - Identification/classification (Q I).** The first step is to secure the electromagnetic counterpart to each LISA source and classify it (DWD or white dwarf with a dark companion). Assuming an at least 1-in-5 identification success rate and 1 hour per candidate spectrum, this implies ~2040 grey/dark moon telescope hours (or ~255 nights). Given this cost, *we advocate for either a dedicated medium-to-large-aperture facility for LISA follow-up or the re-purposing of an existing telescope/instrument for this programme, equipped with an R≈2000 full visible coverage spectrograph*. Such a LISA-focused telescope could serve a very similar purpose to PLATOSpec at La Silla: efficiently filtering candidates from LISA fields. Spectra obtained for candidates that are ultimately unrelated to the LISA signal will still be valuable for broader community science.

**Target II - Phase-resolved spectroscopy (Q II–IV).** This capability could be delivered by a stand-alone instrument or combined with the identification spectrograph described above. Exposure times must be short enough to avoid orbital smearing of spectral features: for a 10-min binary, exposure time should be ≲120 seconds (≤ 0.2 cycles). Current ESO readout times (>20 seconds) dominate overheads, so we strongly advocate for detectors with ***near-zero readout time*** to be hosted by ESO. Because LISA counterparts will be faint and short exposures collect few photons, ***near-zero readout noise detectors*** (current ESO detectors are a few e$^-$/pixel) ***placed on a large-aperture telescope*** is a crucial design requirement as well.

# Full list of supporters

*(Name, Institute, Country)*

Borja Anguiano (Centro de Estudios de Física del Cosmos de Aragón, Spain),
K. G. Arun (Chennai Mathematical Institute, India),
Maria Babiuc Hamilton (Marshall University, USA),
Carles Badenes (University of Pittsburgh, USA),
John G. Baker (Université de Toulouse, France),
Manuel Barrientos (University of Oklahoma, USA),
Enrico Barausse (SISSA, Italy),
Alexander Bobrick (Monash University, Australia),
Elisa Bortolas (INAF – Osservatorio Astronomico di Padova, Italy),
Tamara Bogdanovic (Georgia Institute of Technology, USA),
Alexander Bonilla Rivera (Instituto de Física, Universidade Federal Fluminense, Brazil),
Martin A. Bourne (University of Hertfordshire, United Kingdom),
Warren R. Brown (Center for Astrophysics, Harvard & Smithsonian, USA),
Tomasz Bulik (University of Warsaw, Poland),
Riccardo Buscicchio (University of Milano-Bicocca, Italy),
Mario Cadelano (University of Bologna, Italy),
Ana Caramete (Institute of Space Science – INFLPR Subsidiary, Romania),
Laurentiu-Ioan Caramete (Institute of Space Science – INFLPR Subsidiary, Romania),
Amodio Carleo (INAF – Osservatorio Astronomico di Cagliari, Italy),
Sylvain Chaty (Université Paris Cité, CNRS, Astroparticule et Cosmologie, France),
Michael W. Coughlin (University of Minnesota, USA),
Stefano Covino (INAF / Brera Astronomical Observatory, Italy),
Alexander W. Criswell (Vanderbilt University & Fisk University, USA),
Gergely Dálya (L2IT, Laboratoire des 2 Infinis – Toulouse, Université de Toulouse, France),
Benjamin L. Davis (New York University Abu Dhabi, United Arab Emirates),
Harry Dawson (University of Potsdam, Germany),
Pratika Dayal (Canadian Institute for Theoretical Astrophysics, University of Toronto, Canada),
Kyriakos Destounis (Universidade de Lisboa, Instituto Superior Técnico, Portugal),
Tiziana Di Salvo (University of Palermo, Italy),
Thomas Donlon (University of Alabama in Huntsville, USA),
Matti Dorsch (University of Potsdam, Germany),
Hannah Dykaar (McGill University, Canada),
Kareem El-Badry (California Institute of Technology, USA),
Chris Fryer (Los Alamos National Laboratory, USA),
Ioannis D. Gialamas (Laboratory of High Energy and Computational Physics, Estonia),
Poshak Gandhi (University of Southampton, United Kingdom),
Silvia Gasparotto (CERN, Switzerland),
Sebastien Guillot (IRAP, France),
Leonid Gurvits (Delft University of Technology, The Netherlands),
Konstantinos Nektarios Gourgouliatos (University of Patras, Greece),
Alister W. Graham (Swinburne University of Technology, Australia),
Paul Groot (Radboud University; University of Cape Town & SAAO, Netherlands & South Africa),
Ulrich Heber (Dr. Remeis-Sternwarte, University of Erlangen–Nürnberg, Germany),
Joerg Hennig (University of Otago, New Zealand),
JJ Hermes (Boston University, USA),
Andrei Igoshev (Newcastle University, United Kingdom),
Luca Izzo (INAF – Osservatorio Astronomico di Capodimonte, Italy),
Gaurava K. Jaisawal (DTU Space, Technical University of Denmark, Denmark),
Shang-Jie Jin (University of Western Australia & Northeastern University, Australia & China),

Alexandros Karam (National Institute of Chemical Physics and Biophysics, Estonia),
Nikolaos Karnesis (Aristotle University of Thessaloniki, Greece),
Mark Kennedy (University College Cork, Ireland),
Mukremin Kilic (University of Oklahoma, USA),
Soon-Wook Kim (Korea Astronomy and Space Science Institute, South Korea),
Albert Kong (National Tsing Hua University, Taiwan),
Erika Korb (Astronomical Observatory of Valle d'Aosta, Italy),
Rubina Kotak (University of Turku, Finland),
Kyle Kremer (UC San Diego, USA),
Nina Kunert (Observatoire de la Côte d'Azur, France),
Kristen Lackeos (Deutsches Zentrum für Astrophysik, Germany),
Kunyang Li (Flatiron Institute CCA, USA),
Dan Maoz (Tel Aviv University, Israel),
Filippo Mannucci (INAF – Arcetri, Italy),
Sylvain Marsat (L2IT/CNRS, France),
Lucio Mayer (University of Zurich, Switzerland),
Jonathan Menu (KU Leuven, Belgium),
Hannah Middleton (University of Birmingham, United Kingdom),
M. Coleman Miller (University of Maryland, USA),
Patrick Motl (Indiana University Kokomo, USA),
Samaya Nissanke (University of Potsdam, Germany),
Antonios Nathanail (Academy of Athens, Greece),
Rüdiger Pakmor (Max Planck Institute for Astrophysics, Germany),
Seong Chan Park (Yonsei University, South Korea),
Ferdinando Patat (ESO, Germany),
Alice Perego (Observatoire de la Côte d'Azur, France)
Mauro Pieroni (IEM-CSIC, Spain),
Andrea Possenti (INAF, Italy),
Tom Prince (Caltech, USA),
Max Pritzkuleit (University of Potsdam, Germany),
Varun Pritmani (American Museum of Natural History, USA),
Hans-Walter Rix (Max Planck Institute for Astronomy, Germany),
Elena Maria Rossi (Leiden Observatory, The Netherlands),
Stephan Rossog (University of Hamburg, Germany),
Milton Ruiz (University of Valencia, Spain),
Ashley Ruiter (Australian National University, Australia),
Argyro Sasli (University of Minnesota, USA),
Simone Scaringi (Durham University, United Kingdom),
Raffaella Schneider (Sapienza Università di Roma, Italy),
Mathias Schultheis (Observatoire de la Côte d'Azur, France),
Alberto Sesana  (Università di Milano Bicocca, Italy)
Ken Shen (UC Berkeley, USA),
Nathan Steinle (University of Manitoba, Canada),
Jiří Svoboda (Astronomical Institute of the Czech Academy of Sciences, Czechia),
Arthur Suvorov (University of Tübingen, Germany),
Nicola Tamanini (Laboratoire des 2 Infinis – Toulouse, France),
Thomas M. Tauris (Aalborg University, Denmark),
Ovidiu Tintareanu-Mircea (Institute of Space Science – INFLPR, Romania),
Silvia Toonen (University of Amsterdam, The Netherlands),
Alejandro Torres-Orjuela (BIMSA, China),
Amaury Triaud (University of Birmingham, United Kingdom),

Ruggero Valli (Max Planck Institute for Astrophysics, Germany),
Karthik Mahesh Varadarajan (University of Southampton, United Kingdom),
Alejandro Vigna-Gómez (Niels Bohr Institute, Denmark),
Daniele Vernieri (University of Naples Federico II, Italy),
Digvijay Wadekar (University of Texas at Austin, USA),
Niels Warburton (University College Dublin, Ireland),
Stoytcho Yazadjiev (University of Sofia, Bulgaria),
Shu-Xu Yi (Institute of High Energy Physics, CAS, China),
Weitian Yu (Universität Hamburg, Germany)